# Spin orbit torques and Dzyaloshinskii-Moriya interaction in dual-interfaced Co-Ni multilayers

Jiawei Yu[1†], Xuepeng Qiu[1†], Yang Wu[1], Jungbum Yoon[1], Praveen Deorani[1], Jean Mourad Besbas[1], Aurelien Manchon[2], and Hyunsoo Yang[1]*

[1] Department of Electrical Engineering and Computer Engineering, National University of Singapore, 117576, Singapore

[2] Division of Physical Science and Engineering, King Abdullah University of Science and Technology (KAUST), Thuwal 23955, Saudi Arabia

[†]These authors contributed equally to this work.

Correspondence and requests for materials should be addressed to H.Y. (eleyang@nus.edu.sg)

**We study the spin orbit torque (SOT) and Dzyaloshinskii-Moriya interaction (DMI) in the dual-interfaced Co-Ni perpendicular multilayers. Through the combination of top and bottom layer materials (Pt, Ta, MgO and Cu), SOT and DMI are efficiently manipulated due to an enhancement or cancellation of the top and bottom contributions. However, SOT is found to originate mostly from the bulk of a heavy metal (HM), while DMI is more of interfacial origin. In addition, we find that the direction of the domain wall (DW) motion can be either along or against the electron flow depending on the DW tilting angle when there is a large DMI. Such an abnormal DW motion induces a large assist field required for hysteretic magnetization reversal. Our results provide insight into the role of DMI in SOT driven magnetization switching, and demonstrate the feasibility of achieving desirable SOT and DMI for spintronic devices.**

Current induced domain wall (DW) nucleation, fast DW motion[1-5] and highly efficient magnetization switching[6-8] have recently been reported in HM/FM heterostructures. The underlying mechanism responsible for these observations has been identified to be spin orbit torques (SOTs) which may arise either from the Rashba[6,9-12] and/or spin Hall effect (SHE)[4,5,7]. SOTs have been also explored in exotic systems such as topological insulators[13,14] and a two dimensional electron gas, $LaAlO_3/SrTiO_3$[15]. Electrical currents flowing through heavy metal layers with strong spin orbit coupling, such as Pt and Ta, give rise to a torque on the magnetization of the FM layer, thus affecting the magnetization dynamics and switching the magnetic states. However, an external assist field ($H_{assist}$), parallel or anti-parallel to the electrical current, is necessary for deterministic current induced magnetization switching. A few studies[16-18] have been carried out to explore the role of the assist field in the switching process. It has also been reported that the Dzyaloshinskii–Moriya interaction (DMI)[19,20] plays an important role in forming a Néel-type DW in thin magnetic multilayers, that can be driven efficiently by the SOT arising from the SHE[4,5,21,22]. However, so far there has been little analysis of the DMI effect in the current induced switching process based on direct experimental observation. The correlation between the SOT and DMI lacks exploration.

Recently, large SOTs in Co-Pd multilayers[11] and chiral spin torque in Co-Ni wires[21,23] have been reported, which may lead to a lower switching current density in future memory and logic devices based on SOT. There are several methods presented in literature to manipulate the efficiency of SOT, such as interface doping[12,24], change of the HM or FM material[5,25-28], variation of the HM and FM thicknesses[29-31] and change of annealing conditions [30,32]. In this work, we explore the possibility of combined SOT effect from top and bottom interfaces between HM and FM layers. We investigate the magnitudes of SOTs in MgO (2 nm)/Pt (4

nm)/Co (0.1 nm)/[Ni (0.1 nm)/Co (0.1 nm)]$_4$/X with different capping layers (X = Pt, Ta, MgO, or Cu) as shown in Fig. 1(a). In these systems, there are two HM/FM interfaces, bottom Pt/Co-Ni and top Co-Ni/X, which are utilized to generate SOT. Depending on the choice of capping layer X, one can expect an enhancement or cancellation of these two SOT contributions[26], which provides an interesting prospect for obtaining larger SOT and efficient current-induced magnetization switching through a proper combination of materials in structure engineering. We observe that the net SOT effective field can be tuned by using dual interfaces (top and bottom) with different materials as the capping layer. The SOT strength depends on the relative sign of the SOTs produced by the two interfaces, resulting in a cancellation or enhancement of the net SOT effective fields.

As reported in recent works, the magnitude and sign of DMI can be changed by using different underlayer materials[23,24,33] and thus the induced Néel wall with different chiralities can be driven efficiently in different directions by the SHE[5,21,22]. We observe that the DMI in our devices varies with different capping layers, thereby proving the feasibility of effective DMI engineering by placing two materials with opposite DMI signs on opposite sides of a common FM layer. In addition, we experimentally reveal that a large negative DMI can result in abnormal current induced switching behaviors preventing full hysteretic magnetization reversal. This demonstrates the distinct roles played by the DMI and SOT in the current-driven switching, and indicates pathways towards the improvement of switching properties of these devices.

The Co-Ni multilayer system is utilized in our work because the interfacial perpendicular magnetic anisotropy (PMA) of Co-Ni expedites the structural engineering while maintaining PMA. The Co-Ni multilayer system[34] also provides a low damping and high spin polarization[35,36]. All these properties make the Co-Ni multilayer system a suitable choice to carry out the studies of different capping layers.

**Results**

**SOT measurements.** The SOT induced effective fields are measured by the ac harmonic technique[11,29,32,37] and the measurement schematic is shown in Fig. 1(b). A sinusoidal current ($I_{ac}$) with a magnitude of 10 mA and a frequency of 13.7 Hz is applied to the devices, and two lock-in amplifiers are used to measure the 1$^{st}$ and 2$^{nd}$ harmonic voltages across the Hall bar. The measurement results are shown in Fig. 1(c-f) for different capped devices. The black lines in Fig. 1(c-f) show the 1$^{st}$ harmonic voltage, while the red and blue lines show the result of 2$^{nd}$ harmonic voltages. In the longitudinal measurement data (red line), a positive peak of the 2$^{nd}$ harmonic voltage is observed in the positive field region and a negative peak exists in the negative field region. In the transverse measurement data (blue line), there is a positive peak in the positive and negative field regions. By fitting the 2$^{nd}$ harmonic Hall voltage (see Supplementary Figs. 1 and 2), SOT effective fields are obtained and the results are summarized in Table 1.

The MgO and Cu capped devices are found to have almost identical SOT effective fields, while the Pt capped device shows the smallest effective fields due to the cancellation of SOTs from bottom and top Pt layers. It must be noted that individual contribution to the total SOT from the bottom Pt (4 nm) and capping Pt layer (2 nm) is different, and thus the net SOT is non zero. The device capped with Ta, which has a large spin Hall angle with opposite sign as that of Pt, results in the largest effective field in both longitudinal ($H_L$) and transverse ($H_T$) directions. The extracted torque efficiency, or effective spin Hall angles[11], $\alpha_i = (2e/\hbar) M_S t_F H_i$ ($i$ = L, T) are in the square brackets of Table 1, where $M_S$ is the saturation magnetization and $t_F$ is the thickness of the Co-Ni layer. Similar to the effective fields, the Ta capped device shows the largest effective spin Hall angles. The above experimental results demonstrate the feasibility of tuning SOTs via dual-interfacial structures with different capping layers.

The magnitudes of SOT effective fields agree well with the current induced switching measurements in Fig. 2 (a), which show that the device with Ta capping has the lowest threshold current ($I_{sw}$) for switching (Table 1). The switching phase diagram in Fig. 2(b) shows that $I_{sw}$ gradually decreases as $H_{assist}$ increases in the MgO capped device. Cu or Pt capped devices also show a similar behavior to the MgO capped device (see Supplementary Fig. 4). However, a large longitudinal assist field ($H_{assist} \sim 1000$ Oe) is indispensable for a hysteretic magnetic reversal in the device with Ta capping as shown in Fig. 2(c). Incomplete switching is observed in Ta capped devices with a small assist field (~200 Oe) as shown in Fig. 2 (a). In order to quantify the DMI and its possible role in the observed abnormal switching process, we have carried out DW measurements.

**DMI measurements.** The DMI effective field ($H_{DMI}$) and DMI constant ($D$) are measured by studying the DW behavior in the creep region[38,39]. As indicated in Fig. 3(a), the DW in the film is driven by the out-of-plane component of the applied field, while the in-plane component breaks the rotational symmetry caused by the $H_{DMI}$, facilitating the anisotropic domain expansion. Polar Kerr microscopy is deployed to image the asymmetric DW creep velocity along the direction of an applied in-plane field. The DW creep images in Fig. 3(b) are obtained by overlapping two DW images recorded by Kerr microscopy at different times. The domain distorts along the applied in-plane field with Pt capping, showing that the right hand side edge of the domain moves faster than the left edge. However, domain distorts in a direction opposite to that of applied field in Ta, MgO and Cu capped films, showing that the left edge has a higher creep velocity than the right edge. The opposite asymmetry suggests an inversed orientation of the magnetic moment within the DW, which is illustrated as green arrows in Fig. 3(b). The chirality of DMI stabilizes right-handed Néel walls (↑→↓ or ↓←↑) in the Pt capped film, and left-handed Néel walls (↑←↓ or ↓→↑) in Ta, MgO, and Cu capped devices. It must be noted that the Néel walls in these systems are not always perfect as there is a competition between the longitudinal $H_{DMI}$ and DW anisotropy field[22], of which the former tends to stabilize the Néel-type wall while the latter stabilizes the Bloch-type wall.

The anisotropy field $H_K$ and $M_S$ in the films with 4 different capping layers are measured for DMI calculation. As shown in Table 1, considerable variations of $M_S$ are observed in different capping samples, indicating the presence of magnetically dead layers. To evaluate the impact of the capping layer on the dead layer, we have prepared samples with various Co/Ni thicknesses ($t_{FM}$) for 4 different capping layer films. The magnetization per unit area as a function of $t_{FM}$ is plotted in Fig. 4. The magnetically dead layer thickness ($t_{DL}$) is extracted and summarized in Table 1.

The DW creep velocity under the influence of an external in-plane field is measured as shown in Fig. 3(c). The value of $H_{DMI}$ is obtained by fitting the data with a model (see Methods). The effective DMI constant ($D$) is extracted by using $D = \mu_0 H_{DMI} M_S \Delta$. Notably, the DMI strength in the Ta capped case ($H_{DMI}$ = -1038.6 Oe, $D$ = -0.394 mJ/m$^2$) is much larger than that in the Pt, MgO, and Cu capped films (Table 1), which is large enough to stabilize left-hand Néel walls.

**Discussion**

**Dual-interfacial manipulation of SOT and DMI.** Recently, room temperature magnetic skyrmions have attracted significant research interests. Dual-interfaced systems are preferred for skyrmions experiments[40-42] due to their large effective DMI. A fine understanding of the different contributions of the top and bottom interfaces to the SOT and DMI is crucial to such achievements. We estimate $H_L$ and $H_T$ (as well as $\alpha_L$ and $\alpha_T$) from a single (top or bottom) HM/FM interface by assuming that $H_L$ and $H_T$ from top and bottom interfaces simply add up algebraically. It should be noted that we consider the current shunting effect in the calculation of SOT effective fields, and we normalize $H_{L/T}$ in each HM layer with their own current density. Detailed discussion of the current shunting calculation can be found in Supplementary Information S8. $H_L$, $H_T$, and DMI of the bottom Pt/FM interface are extracted from the MgO capped device, since MgO has negligible spin orbit coupling and there is no current in the MgO capping layer. The results are summarized in Table 2. The small values of $H_L$ and $H_T$ from the top FM/Cu interface are within the measurement sensitivity. The extracted values are $H_L$ = 116.7 Oe (per $10^8$ A/cm$^2$) and $H_T$ = 125.6 Oe (per $10^8$ A/cm$^2$) from top FM/Pt interface, and $H_L$ = -2502.2 Oe (per $10^8$ A/cm$^2$) and $H_T$ = -4213.8 Oe (per $10^8$ A/cm$^2$) from top FM/Ta interface. The ratios $|H_L/H_T|$ are 1.2, 0.929, and 0.594 from bottom Pt/FM, top FM/Pt, and top FM/Ta interfaces, respectively. This result is in contrast with other recent studies which report a larger value of $|H_L/H_T|$ in Pt[5] and a smaller value of $|H_L/H_T|$ in Ta[29,37] compared to our values. The DMI constants from bottom Pt/FM, top FM/Pt, and top FM/Ta interfaces are also calculated using a similar approach.

Three important implications can be drawn from the top/bottom interface contribution to SOT and DMI presented in Table 2. First, by comparing the SOT data from the 1$^{st}$ and 3$^{rd}$ column, it is clear that the bottom Pt/Co interface contributes more to SOTs in the Pt/FM/Pt structure. Considering that the bottom Pt (4 nm) is thicker than the top Pt (2 nm), we can conclude that the bulk contribution from Pt is more important to SOTs in our system. Second, the DMI constant from the top FM/Pt interface (column 3) is about five times larger than that from the bottom Pt/FM interface (column 1), suggesting that the top FM/Pt interface dominates for DMI in the Pt/FM/Pt structure, even though the top Pt (2 nm) is thinner than the bottom Pt (4 nm). This gives an important clue that the DMI may be more closely related to the interfacial contributions. More importantly, the above results indicate that the SOT and DMI can be tuned separately by using different capping materials and their thickness. Third, the top FM/Ta and top FM/Pt interfaces show DMI constants with similar magnitudes but opposite signs. Our findings are different from some recent works[43,44], which show that the Pt/FM and Ta/FM interfaces give DMI constants with the same sign, and the Pt/FM interface gives much larger DMI constants compared to Ta/FM interface. The different DMI signs of Pt and Ta can be utilized for future large DMI engineering. Our work shows the feasibility of DMI enhancement by placing two materials with opposite DMI signs on a common FM layer. This capping layer engineering opens a window for structural engineering of DMI in ferromagnetic material systems, which can be utilized for future skyrmions studies.

**Role of DMI in SOT current induced magnetization switching.** We discuss the role of DMI in SOT induced magnetization switching based on experimental observations. It was recently proposed[16] that in nucleation driven magnetization reversal of inversion asymmetric heterostructures, the DMI field must be overcome by a large external magnetic field in order to enable spin Hall driven expansion of the nucleated domains in all directions. While such a scenario qualitatively explains the switching behaviors of Cu, MgO, and Pt capped systems, it fails to account for the case of Ta capped system. Figure 5(a) and (c) show a normal and abnormal SOT current induced switching in Cu (small DMI and small $H_{assist}$) and Ta capped (large DMI is not compensated by the small $H_{assist}$) devices with a 400 Oe assist field, respectively. The switching in the Ta capped device with a 1000 Oe assist field (large DMI is compensated by the large $H_{assist}$) is presented in Fig. 5(e). For detailed analysis, the switching process is divided into several detailed steps marked with numbers from 1 to 5. Each switching step is imaged by Kerr microscopy as shown in Fig. 5(b), 5(d), and 5(f) for the Cu ($H_{assist}$ = 400 Oe), Ta (400 Oe), and Ta (1000 Oe) cases, respectively.

Now we focus on the interesting case [Fig. 5(c)] that shows the anhysteretic SOT switching, and we will explain the role of DMI in the SOT switching process in details. Figure 5(d) shows that the magnetization reversal occurs through propagation of a tilted DW along the electron flow (panels 1 and 2) until it reaches a strong pinning site at the exit of the Hall cross (panel 3). At this stage, the DW tilting almost vanishes and its depinning takes place against the electron flow (panel 4). The change of sign of the DW velocity between panel 3 and 4 holds the key to understand the absence of hysteresis for this system as seen in Fig. 5(c), and indicates that a strong connection exists between the DW tilting and its direction of motion.

We first attempt to utilize previous understanding of chiral spin torque driven DW motion[4,5,16,21,22] in order to explain the DW motion along the current direction shown in the Ta capped case [panel 3-5 in Fig. 5(d)] as follows. When a DW is created in the Ta capped magnetic wire [panel 3 of Fig. 5(d)], the negative DMI constant in Ta capped device tends to stabilize a left-hand Néel wall ($m_x < 0$) as shown in Fig. 6(a). The longitudinal SOT effective field [$H_L$, shown as small red arrows in Fig. 6(a)] has a component along the +z direction and thus favors the up-alignment of the magnetic moments. This drives the DW along the current flow[5] and gives rise to the abnormal backward magnetization reversal, as shown in Fig. 5(d). However, this explanation cannot cover the whole switching process of the Ta capped device.

The velocity of a tilted magnetic DW can be written as (see Supplementary section 6)[45] $\partial_t q \approx -(\gamma\pi\Delta/2\alpha)H_L(\sin\psi/\cos\chi)$, where $\psi$ is the azimuthal angle of the magnetization in the DW and $\chi$ is the tilting angle of the wall with respect to the transverse direction as indicated in Fig. 6(b). The non-adiabatic torque is neglected due to the ultrathin thickness of the magnetic layer. Figure 6(c) displays the calculated domain wall velocity as a function of the domain wall tilting for the case of Ta capping at $H_{assist}$ = 400 Oe (black symbols) and $H_{assist}$ = 1000 Oe (red symbols). The azimuthal angle of the magnetic moment in the center of the wall is given in the inset. It appears that the azimuthal angle $\psi$ (see inset) changes sign when the tilting angle changes in the case of $H_{assist}$ = 400 Oe, while it remains positive in the case of $H_{assist}$ = 1000 Oe. As a result, the velocity remains negative (i.e. with the electron flow) in the case of $H_{assist}$ = 1000 Oe, while it changes sign when $H_{assist}$ = 400 Oe.

Hence, we can explain our data as follows. In the case shown in Fig. 5(d), the tilting angle ($\chi$) is large when the DW lies in the middle of the wire [panel 1 in Fig. 5(d)] and the combination of the different fields results in $\sin\psi>0$. Hence, the DW moves along the electron flow ($\partial_t q<0$). This situation corresponds to region 1 (blue shade) in Fig. 6(c). When the tilting of the DW is reduced due to a strong symmetric pinning [panel 3 in Fig. 5(d)], the azimuthal angle ($\psi$) becomes negative ($\sin\psi<0$) and the DW moves against the electron flow. This situation corresponds to region 2 (red shade) in Fig. 6(c). This change of sign of the velocity prevents hysteretic switching to occur. In contrast, when the DMI is compensated by a large in-plane field [Fig. 5(a,b,e,f)], the tilting angle is intermediate and relatively constant. The azimuthal angle remains small but positive, therefore, the DW always moves along the electron flow, independent of the DW tilting (region 3, green shade). The corresponding magnetization configurations are represented in the images with the same panel number in Fig. 5(g,h). The transition from panels 1 to 4 in Fig. 5(d) implies that the domain wall velocity changes its sign. Therefore, there should be a region of parameters where the domain wall velocity vanishes. Close to this transition point, our rigid model does not properly apply as thermally activated domain wall nucleation (e.g. at the edges of the wire) tends to dominate over domain wall propagation. Nonetheless, the phenomenological model explains qualitatively the observation in the domain wall propagation regime.

In conclusion, we show the feasibility of structural engineering in SOT based devices, and attain a large SOT effective field and a large DMI in the Pt/Co-Ni/Ta system because of the different signs of SOT and DMI induced by the top and bottom heavy metal materials. Similarly, we observe a small SOT and DMI in the Pt/Co-Ni/Pt system because of the cancellation of SOT and DMI from the two heavy metal layers. By extracting the top and bottom HM/FM interfacial contributions to SOT and DMI, we find that the bulk contributions are important in SOT while interfacial contributions are dominant in DMI. We show that the SOT and DMI do not necessarily originate from the same interface, and can be engineered separately. The mechanism of SOT current induced switching is also explored and we demonstrate that a large in-plane assist field is necessary (~1000 Oe) for full switching in the Ta capped sample due to a large negative DMI (-1038.6 Oe). In contrast, systems with smaller DMI such as the Cu capped sample only need a moderate assist field to achieve hysteretic reversal. Our findings shed light on the structural engineering in SOT based devices and the role of DMI in the current induced SOT switching.

**Methods**

**Sample preparation.** The film stacks used in this experiment are (thickness unit in nm): substrate/MgO (2)/Pt (4)/Co (0.1)/[Ni (0.1)/Co (0.1)]$_4$/X, where X = Pt (2), Ta (2)/MgO (2)/Al$_2$O$_3$ (1), MgO (2)/Al$_2$O$_3$ (1), or Cu (2)/Al$_2$O$_3$ (1). The samples were deposited on a thermally oxidized silicon wafer by ultra-high vacuum magnetron sputtering at room temperature. Ar (~2.3 mTorr) gas was used during the sputtering process. The films were then patterned into cross-shaped wires with the width of 10 μm by photolithography and ion milling processes.

**DW velocity measurement.** A circular domain is created on the as-deposited film surface around a nucleation center. An in-plane field $H$ (~ 0 – 1000 Oe) is applied to the film with a small out-of-plane tilting angle (θ ~ 8 - 10°). Polar Kerr microscopy is used to monitor the DW motion. The DW velocity with the presence of an applied in-plane field ($H$) can be determined from the DW displacement and the interval between two Kerr images. Each data point in Fig. 3(c) is an average of five measurements to reduce the measurement error. The anisotropy field $H_K$ and saturation magnetization $M_S$ in 4 different capping layers (Table 1) were obtained from vibrating sample magnetometer (VSM) measurements (see Supplementary Fig. 6).

**Fitting of DW velocity curve and obtain $H_{DMI}$.** The DW velocity under an out-of-plane applied field $H_z$ in the creep region can be described by the creep law[46] $v = v_0 \exp(-\zeta H_z^{-1/4})$, where $v_0$ is the characteristic speed fitting parameter and $\zeta$ can be expressed as $\zeta = \zeta_0 \left[\sigma(H_x)/\sigma_0\right]^{1/4}$. $\zeta_0$ is the scaling fitting parameter, and $\sigma$ is the DW energy density as a function of $H_x$[38,39]:

$$\sigma(H_x) = \begin{cases} \sigma_0 - \dfrac{\pi^2 \Delta M_S^2}{8K_D}(\mu_0 H_x + \mu_0 H_{\text{DMI}})^2 & \text{for } |\mu_0 H_x + \mu_0 H_{\text{DMI}}| < \dfrac{4K_D}{\pi M_S}, \\ \sigma_0 + 2K_D \Delta - \pi \Delta M_S |\mu_0 H_x + \mu_0 H_{\text{DMI}}| & \text{otherwise,} \end{cases} \quad (1)$$

where $\sigma_0 = 2\pi\sqrt{AK_{0,\text{eff}}}$ is the Bloch wall energy density, $\Delta = \sqrt{A/K_{0,\text{eff}}}$ is the DW width, $K_{0,\text{eff}} = \mu_0 H_K M_S/2$ is the effective magnetic anisotropy constant, and $4K_D/(\pi M_S)$ is the magnetic field required to change the DW into Néel type. $K_D = N_x \mu_0 M_S^2/2$ is the DW anisotropy energy density which represents the magnetostatic energy difference between the Bloch and Néel DW[47]. Demagnetization coefficient ($N_x$) of 0.22 and exchange stiffness constant (*A*) of 16 pJ/m are used.

**References**

1. Moore, T. A. *et al.* High domain wall velocities induced by current in ultrathin Pt/Co/AlOx wires with perpendicular magnetic anisotropy. *Appl. Phys. Lett.* **93**, 262504 (2008).
2. Miron, I. M. *et al.* Fast current-induced domain-wall motion controlled by the Rashba effect. *Nat. Mater.* **10**, 419-423 (2011).
3. Chiba, D. *et al.* Electric-field control of magnetic domain-wall velocity in ultrathin cobalt with perpendicular magnetization. *Nat. Commun.* **3**, 888 (2012).
4. Haazen, P. P. J. *et al.* Domain wall depinning governed by the spin Hall effect. *Nat. Mater.* **12**, 299-303 (2013).
5. Emori, S., Bauer, U., Ahn, S.-M., Martinez, E. & Beach, G. S. D. Current-driven dynamics of chiral ferromagnetic domain walls. *Nat. Mater.* **12**, 611-616 (2013).
6. Miron, I. M. *et al.* Perpendicular switching of a single ferromagnetic layer induced by in-plane current injection. *Nature* **476**, 189-193 (2011).
7. Liu, L., Lee, O. J., Gudmundsen, T. J., Ralph, D. C. & Buhrman, R. A. Current-Induced Switching of Perpendicularly Magnetized Magnetic Layers Using Spin Torque from the Spin Hall Effect. *Phys. Rev. Lett.* **109**, 096602 (2012).
8. Liu, L. *et al.* Spin-Torque Switching with the Giant Spin Hall Effect of Tantalum. *Science* **336**, 555-558 (2012).
9. Miron, I. M. *et al.* Current-driven spin torque induced by the Rashba effect in a ferromagnetic metal layer. *Nat. Mater.* **9**, 230-234 (2010).
10. Kim, K.-W., Seo, S.-M., Ryu, J., Lee, K.-J. & Lee, H.-W. Magnetization dynamics induced by in-plane currents in ultrathin magnetic nanostructures with Rashba spin-orbit coupling. *Phys. Rev. B* **85**, 180404 (2012).
11. Jamali, M. *et al.* Spin-Orbit Torques in Co/Pd Multilayer Nanowires. *Phys. Rev. Lett.* **111**, 246602 (2013).
12. Qiu, X. *et al.* Spin–orbit-torque engineering via oxygen manipulation. *Nat. Nanotechnol.* **10**, 333-338 (2015).
13. Mellnik, A. R. *et al.* Spin-transfer torque generated by a topological insulator. *Nature* **511**, 449-451 (2014).
14. Wang, Y. *et al.* Topological Surface States Originated Spin-Orbit Torques in $Bi_2Se_3$. *Phys. Rev. Lett.* **114**, 257202 (2015).
15. Narayanapillai, K. *et al.* Current-driven spin orbit field in $LaAlO_3/SrTiO_3$ heterostructures. *Appl. Phys. Lett.* **105**, 162405 (2014).
16. Lee, O. J. *et al.* Central role of domain wall depinning for perpendicular magnetization switching driven by spin torque from the spin Hall effect. *Phys. Rev. B* **89**, 024418 (2014).
17. Lo Conte, R. *et al.* Spin-orbit torque-driven magnetization switching and thermal effects studied in Ta\CoFeB\MgO nanowires. *Appl. Phys. Lett.* **105**, 122404 (2014).
18. Legrand, W., Ramaswamy, R., Mishra, R. & Yang, H. Coherent Subnanosecond Switching of Perpendicular Magnetization by the Fieldlike Spin-Orbit Torque without an External Magnetic Field. *Phys. Rev. Appl* **3**, 064012 (2015).
19. Moriya, T. New Mechanism of Anisotropic Superexchange Interaction. *Phys. Rev. Lett.* **4**, 228 (1960).
20. Fert, A., Cros, V. & Sampaio, J. Skyrmions on the track. *Nat. Nanotechnol.* **8**, 152-156 (2013).

21   Ryu, K.-S., Thomas, L., Yang, S.-H. & Parkin, S. Chiral spin torque at magnetic domain walls. *Nat. Nanotechnol.* **8**, 527-533 (2013).
22   Franken, J. H., Herps, M., Swagten, H. J. M. & Koopmans, B. Tunable chiral spin texture in magnetic domain-walls. *Sci. Rep.* **4**, 5248 (2014).
23   Ryu, K.-S., Yang, S.-H., Thomas, L. & Parkin, S. S. P. Chiral spin torque arising from proximity-induced magnetization. *Nat. Commun.* **5**, 3910 (2014).
24   Torrejon, J. *et al.* Interface control of the magnetic chirality in CoFeB/MgO heterostructures with heavy-metal underlayers. *Nat. Commun.* **5**, 4655 (2014).
25   Pai, C.-F. *et al.* Spin transfer torque devices utilizing the giant spin Hall effect of tungsten. *Appl. Phys. Lett.* **101**, 122404 (2012).
26   Woo, S., Mann, M., Tan, A. J., Caretta, L. & Beach, G. S. D. Enhanced spin-orbit torques in Pt/Co/Ta heterostructures. *Appl. Phys. Lett.* **105**, 212404 (2014).
27   Pai, C.-F. *et al.* Enhancement of perpendicular magnetic anisotropy and transmission of spin-Hall-effect-induced spin currents by a Hf spacer layer in W/Hf/CoFeB/MgO layer structures. *Appl. Phys. Lett.* **104**, 082407 (2014).
28   Loong, L. M., Deorani, P., Qiu, X. & Yang, H. Investigating and engineering spin-orbit torques in heavy metal/Co$_2$FeAl$_{0.5}$Si$_{0.5}$/MgO thin film structures. *Appl. Phys. Lett.* **107**, 022405 (2015).
29   Kim, J. *et al.* Layer thickness dependence of the current-induced effective field vector in Ta|CoFeB|MgO. *Nat. Mater.* **12**, 240-245 (2013).
30   Wang, Y., Deorani, P., Qiu, X., Kwon, J. H. & Yang, H. Determination of intrinsic spin Hall angle in Pt. *Appl. Phys. Lett.* **105**, 152412 (2014).
31   Fan, X. *et al.* Quantifying interface and bulk contributions to spin–orbit torque in magnetic bilayers. *Nat. Commun.* **5**, 3042 (2014).
32   Garello, K. *et al.* Symmetry and magnitude of spin-orbit torques in ferromagnetic heterostructures. *Nat. Nanotechnol.* **8**, 587-593 (2013).
33   Chen, G. *et al.* Tailoring the chirality of magnetic domain walls by interface engineering. *Nat. Commun.* **4**, 2671 (2013).
34   Daalderop, G. H. O., Kelly, P. J. & den Broeder, F. J. A. Prediction and confirmation of perpendicular magnetic anisotropy in Co/Ni multilayers. *Phys. Rev. Lett.* **68**, 682 (1992).
35   Shigemi, M. *et al.* Gilbert Damping in Ni/Co Multilayer Films Exhibiting Large Perpendicular Anisotropy. *Appl. Phys. Exp.* **4**, 013005 (2011).
36   Song, H.-S. *et al.* Observation of the intrinsic Gilbert damping constant in Co/Ni multilayers independent of the stack number with perpendicular anisotropy. *Appl. Phys. Lett.* **102**, 102401 (2013).
37   Qiu, X. *et al.* Angular and temperature dependence of current induced spin-orbit effective fields in Ta/CoFeB/MgO nanowires. *Sci. Rep.* **4**, 4491 (2014).
38   Je, S.-G. *et al.* Asymmetric magnetic domain-wall motion by the Dzyaloshinskii-Moriya interaction. *Phys. Rev. B* **88**, 214401 (2013).
39   Hrabec, A. *et al.* Measuring and tailoring the Dzyaloshinskii-Moriya interaction in perpendicularly magnetized thin films. *Phys. Rev. B* **90**, 020402 (2014).
40   Dupe, B., Bihlmayer, G., Bottcher, M., Blugel, S. & Heinze, S. Engineering skyrmions in transition-metal multilayers for spintronics. *Nat. Commun.* **7**, 11779 (2016).
41   Woo, S. *et al.* Observation of room-temperature magnetic skyrmions and their current-driven dynamics in ultrathin metallic ferromagnets. *Nat. Mater.* **15**, 501-506 (2016).

42    Moreau Luchaire, C. *et al.* Additive interfacial chiral interaction in multilayers for stabilization of small individual skyrmions at room temperature. *Nat. Nanotechnol.* **11**, 444-448 (2016).
43    Perez, N. *et al.* Chiral magnetization textures stabilized by the Dzyaloshinskii-Moriya interaction during spin-orbit torque switching. *Appl. Phys. Lett.* **104**, 092403 (2014).
44    Emori, S. *et al.* Spin Hall torque magnetometry of Dzyaloshinskii domain walls. *Phys. Rev. B* **90**, 184427 (2014).
45    Boulle, O. *et al.* Domain Wall Tilting in the Presence of the Dzyaloshinskii-Moriya Interaction in Out-of-Plane Magnetized Magnetic Nanotracks. *Phys. Rev. Lett.* **111**, 217203 (2013).
46    Gorchon, J. *et al.* Pinning-Dependent Field-Driven Domain Wall Dynamics and Thermal Scaling in an Ultrathin Pt/Co/Pt Magnetic Film. *Phys. Rev. Lett.* **113**, 027205 (2014).
47    Jung, S.-W., Kim, W., Lee, T.-D., Lee, K.-J. & Lee, H.-W. Current-induced domain wall motion in a nanowire with perpendicular magnetic anisotropy. *Appl. Phys. Lett.* **92**, 202508 (2008).

**Acknowledgments**

This research was supported by the National Research Foundation (NRF), Prime Minister's Office, Singapore, under its Competitive Research Programme (CRP award no. NRFCRP12-2013-01). A.M. was supported by the King Abdullah University of Science and Technology (KAUST). H.Y. is a member of the Singapore Spintronics Consortium (SG-SPIN).

**Author contributions**

J.Yu, X.Q., A.M. and H.Y. planned the project. J.Yu and X.Q. fabricated the devices. J.Yu, X.Q., Y.W., J.Yoon and J.M.B. performed measurements. J.Yu, X.Q. and P.D. analyzed the data with the help of A.M. and H.Y. A.M. did modeling. All authors discussed the results and commented on the manuscript. J.Yu, X.Q., P.D., A.M. and H.Y. wrote the manuscript. H.Y. supervised the project.

**Additional information**

Supplementary information accompanies this paper at http://www.nature.com/Scientificreports

Competing financial interests: The authors declare no competing financial interests.

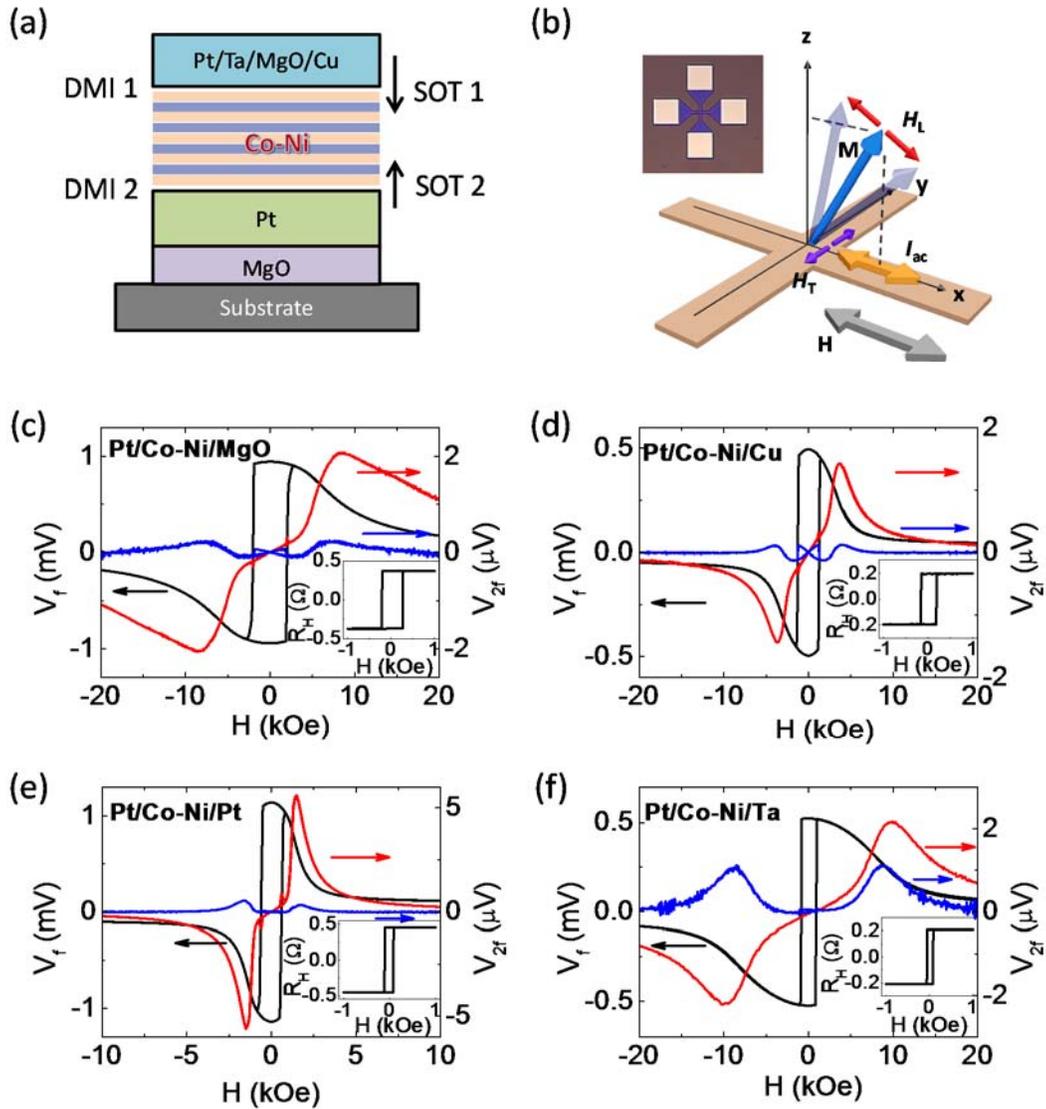

**Figure 1 | Device geometry and SOT measurements.** (a) Schematic of dual-interfacial engineering with different top capping layers, with SOT and DMI arising from both top and bottom interfaces. (b) The measurement schematic for SOT effective fields. Blue arrows present the magnetization directions; gray arrows show the applied field directions and orange arrows give the ac current directions along the ±$x$-direction. Inset: Optical microscope image of the fabricated device. Both the channel and Hall bar widths are 10 μm. (c-f) First and second harmonic voltages in MgO (c), Cu (d), Pt (e) and Ta (f) capped devices. Black lines present the first harmonic voltages. The longitudinal second harmonic voltage is presented as the red line, while the transverse one is shown as the blue line. Each inset shows the out-of-plane anomalous Hall loop.

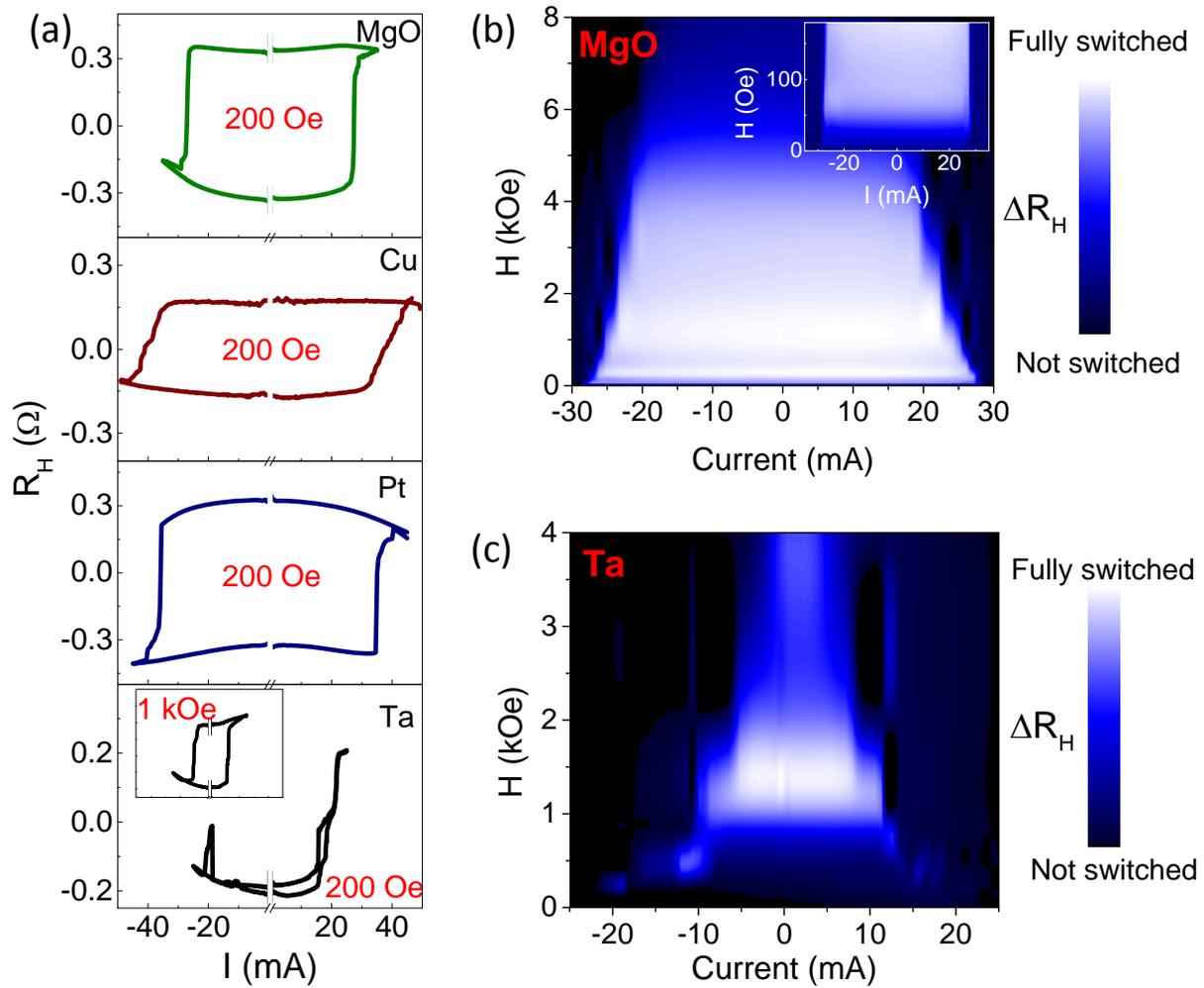

**Figure 2 | SOT induced magnetization switching.** (a) The SOT induced magnetization switching loops with a 200 Oe assist field ($H_{assist}$) in the four different capping layer devices. Inset: the current induced switching loop in the Ta case with $H_{assist}$ = 1 kOe. The inset figure shares the same scale with the main figure. (b) The phase diagram of the SOT induced switching with different $H_{assist}$ of the MgO capped device. Inset: magnification of the main plot in the low field region. (c) The phase diagrams of the Ta capped device.

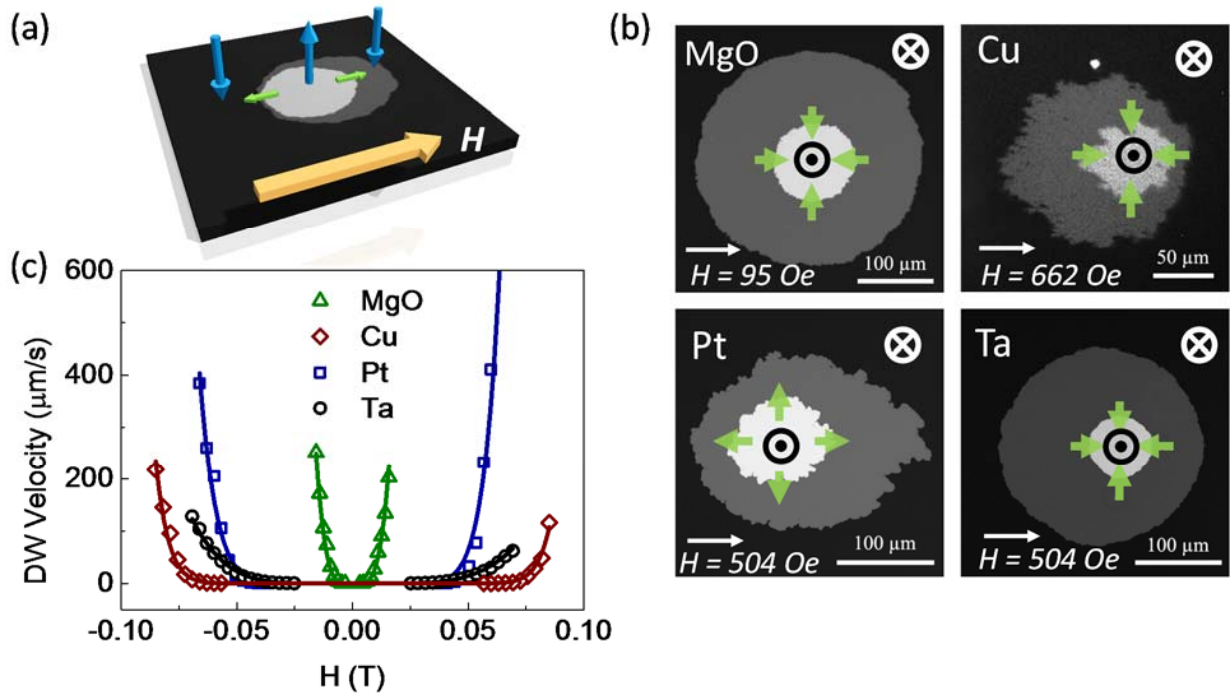

**Figure 3 | DMI measurements.** (a) Illustration of the Kerr microscopy measurement to quantify the DMI effective field ($H_{DMI}$). The magnetization directions of the domains are shown with the blue arrows, and the $H_{DMI}$ are indicated by green arrows. The thick yellow arrow indicates the applied in-plane field. (b) The anisotropic domain wall expansion with an in-plane magnetic field. The magnetic field strength and direction are indicated at the bottom of each picture. Green arrows in each graph indicate the equilibrium magnetization direction within the DW. (c) The asymmetric domain wall creep velocity as a function of the applied in-plane magnetic field for different capping layer films. The symbols are the experimental results and the solid lines show the fitting curves.

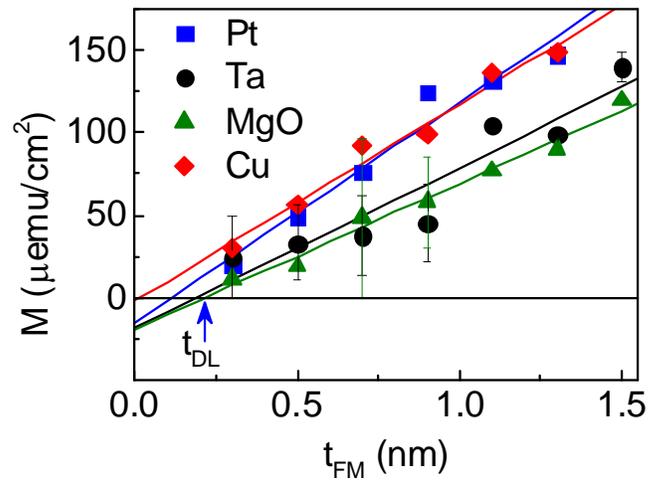

**Figure 4 | Magnetization per unit area versus the thickness of Co/Ni layer in four different capping cases.** The symbols are the experimental results and the solid lines show the linear fittings. The dead layer thickness ($t_{DL}$) is shown as the x-axis intercept.

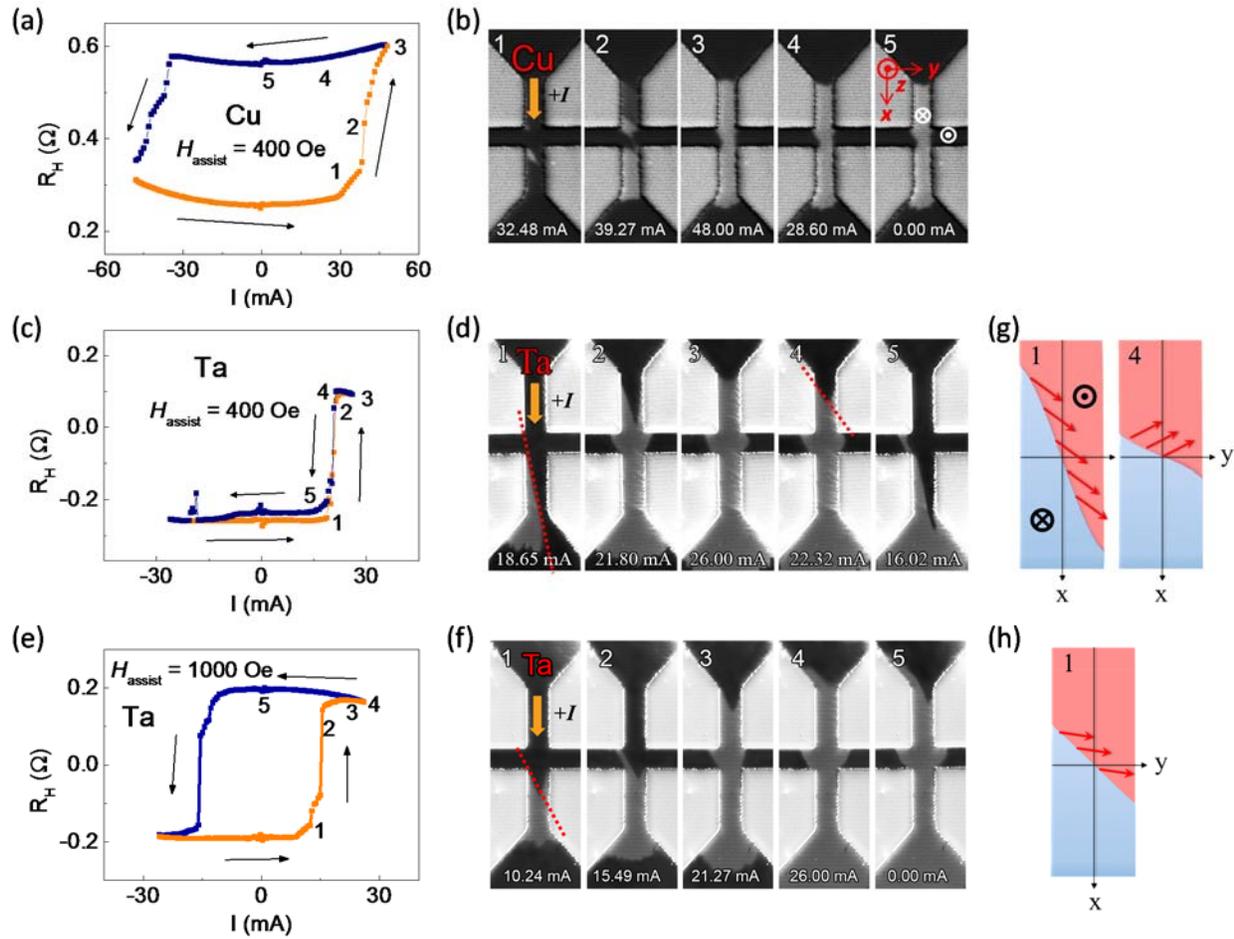

**Figure 5 | Hall measurements and Kerr images of SOT current induced magnetization switching.** (a) The SOT current induced switching curve in the Cu capped device with a 400 Oe assist field. (b) The DW configurations captured by Kerr microscope simultaneously with the SOT current induced switching measurements. The sample is pre-magnetized as magnetization pointing out of plane (dark part in the Kerr image) and the light parts show the switched region (magnetization pointing into the plane). The magnitude of applied current is indicated at the bottom. The switching curve and Kerr images in the Ta capped device are displayed in (c) and (d), respectively, with a 400 Oe assist field. (e,f) The switching curve and Kerr images of the Ta capped device with a 1000 Oe assist field. (g) Domain wall and magnetic moment configurations. Panel 1: strong tilting and positive azimuthal angle, corresponding to panel 1 of Fig. 5(d). Panel 4: small tilting and negative azimuthal angle, corresponding to panel 4 of Fig. 5(d). (h) Intermediate tilting with a compensated DMI by a strong assist field, corresponding to panel 1 of Fig. 5(f).

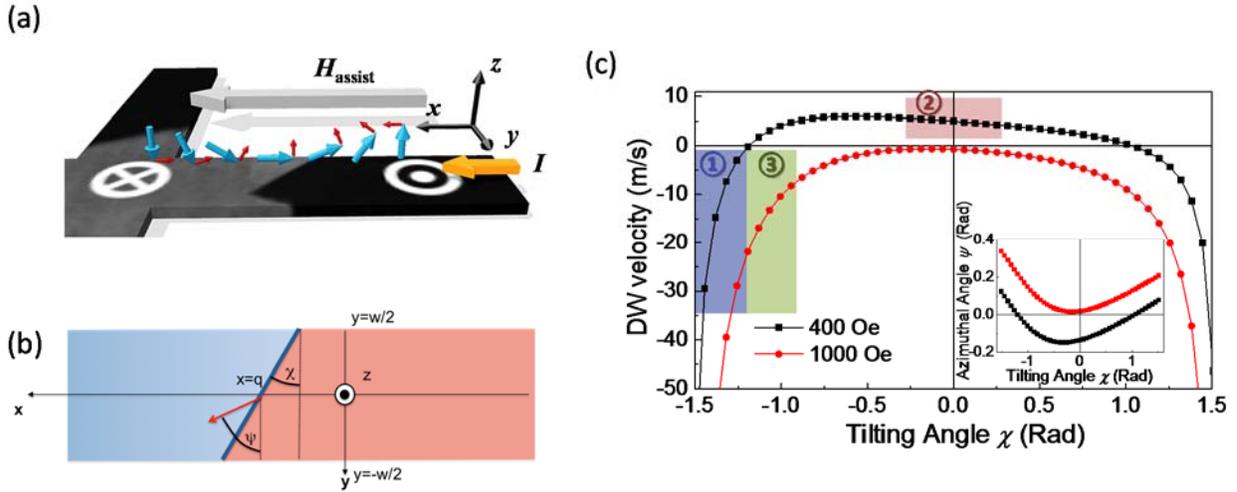

**Figure 6 | Theoritical analysis of abnormal backward magnetization reversal.** (a) The schematic diagram of the DMI stabilized left-hand Néel wall in Ta capped devices and the corresponding longitudinal SOT effective field ($H_L$) directions. The thick blue arrows show the magnetization directions in different regions and red arrows indicate the $H_L$. The in-plane assist field ($H_{assist}$) is presented in the light gray arrow and the injected current is displayed as the thick orange arrow. (b) Schematics of the tilted magnetic domain wall. The red and blue regions correspond to oppositely magnetized domains, and the magnetic wall located as the position $x = q$ is tilted with an angle $\chi$ with respect to the transverse direction $y$. (c) Domain wall velocity as a function of the tilting angle. The shaded regions correspond to the experimentally relevant cases for Ta: $H_{assist}$ = 400 Oe (blue region-large tilting; red region-small tilting) and $H_{assist}$ = 1000 Oe (green region-intermediate tilting).

| Table 1 \| Summary of experimental results of SOT and DMI. | | | | |
|---|---|---|---|---|
| **Structures** | Pt/Co-Ni/MgO | Pt/Co-Ni/Cu | Pt/Co-Ni/Pt | Pt/Co-Ni/Ta |
| $H_L$ (Oe per $10^8$ A/cm$^2$) [$\alpha_L$] | 272.4 [0.043] | 244.7 [0.087] | 138.4 [0.046] | 1180.7 [0.291] |
| $H_T$ (Oe per $10^8$ A/cm$^2$) [$\alpha_T$] | 227.0 [0.036] | 220.3 [0.078] | 79.2 [0.026] | 1700.2 [0.418] |
| $H_L/H_T$ | 1.20 | 1.11 | 1.75 | 0.69 |
| $I_{sw}$ (mA) | 27.3 (200 Oe) | 38.1 (200 Oe) | 35.5 (200 Oe) | 18.9 (200 Oe) |
| $H_{DMI}$ (Oe) | -155.9 | -192.3 | 348.8 | -1038.6 |
| $D$ (mJ/m$^2$) | -0.053 | -0.117 | 0.195 | -0.394 |
| $M_S$ (emu/cc) | 612.68 | 1150.2 | 1075.5 | 749.47 |
| $H_K$ (Oe) | 8500 | 5000 | 5500 | 10000 |
| $t_{DL}$ (nm) | 0.214 | 0.018 | 0.111 | 0.176 |

| Table 2 \| Contributions of different interfaces to SOT and DMI. | | | | |
|---|---|---|---|---|
| **Interface** | Bottom Pt | Top Cu | Top Pt | Top Ta |
| $H_L$ (Oe per $10^8$A/cm$^2$) [$\alpha_L$] | 225.2 [0.061] | -5.9 [-0.002] | 116.7 [0.038] | -2502.2 [-0.616] |
| $H_T$ (Oe per $10^8$A/cm$^2$) [$\alpha_T$] | 187.6 [0.051] | -14.0 [-0.005] | 125.6 [0.041] | -4213.8 [-1.037] |
| $H_L/H_T$ | 1.200 | 0.419 | 0.929 | 0.594 |
| $D$ (mJ/m$^2$) | -0.053 | 0.064 | -0.248 | 0.341 |